\documentclass[12pt]{article}
\usepackage{amsfonts}

\begin{document}

\title{Quantum games and social norms. The quantum ultimatum game}
\author{R. Vilela Mendes\thanks{%
Grupo de F\'{\i }sica Matem\'{a}tica, Complexo Interdisciplinar,
Universidade de Lisboa, Av. Gama Pinto 2, 1699 Lisboa Codex, Portugal} 
\thanks{%
e-mail: vilela@cii.fc.ul.pt}}
\date{}
\maketitle

\begin{abstract}
The noncooperative Nash equilibrium solution of classical games corresponds
to a rational expectations attitude on the part of the players. However, in
many cases, games played by human players have outcomes very different from
Nash equilibria.

A restricted version of quantum games is proposed to implement, in
mathematical games, the interplay of self-interest and internalized social
norms that rules human behavior.
\end{abstract}


Classical game theory is a mathematical framework for decision problems with
rational rules and rational players. It is used in many situations in
economics, social sciences and communication. An important item, leading to
a solution, is the notion of noncooperative \textit{Nash equilibrium} $x^{*}$%
, 
\[
x^{*}Px^{*}\geq xPx^{*} 
\]
$P$ being the payoff matrix and $x$ and $x^{*}$ mixed strategy vectors. It
means that $x^{*}$ is an ideal strategy when played against itself or that
no player can improve his payoff by changing his strategy, when the
strategies of the other players are fixed.

The Nash equilibrium solutions correspond to the purely self-interested
attitude where, given any environment situation, each player tries to
maximize his gains regardless of what happens to the other players. It is
the rational expectations attitude of what has been called the \textit{Homo
oeconomicus,} a notion which is at the basis of many theoretical economics
constructs. It is therefore important to check the applicability of such
notion in human societies. Experiments have been carried out and the problem
is that in many cases, when played by human players, games have outcomes
very different from the Nash equilibrium points. An interesting case is the 
\textit{ultimatum game}\cite{Guth}. A simplified version of this game is the
following:

One of the players (the \textit{proposer}) receives 100 coins which he is
told to divide into two non-zero parts, one for himself and the other for
the other player (the \textit{responder}). If the responder accepts the
split, it is implemented. If the responder refuses, nothing is given to the
players. Consider, for example, a simple payoff matrix corresponding to two
different proposer offers 
\begin{equation}
\begin{tabular}{|c|c|c|}
\hline
& $R_{0}$ & $R_{1}$ \\ \hline
$P_{0}$ & $
\begin{array}{l}
\left| 00\right\rangle \\ 
a,c
\end{array}
$ & $
\begin{array}{l}
\left| 01\right\rangle \\ 
0,0
\end{array}
$ \\ \hline
$P_{1}$ & $
\begin{array}{l}
\left| 10\right\rangle \\ 
b,b
\end{array}
$ & $
\begin{array}{l}
\left| 11\right\rangle \\ 
0,0
\end{array}
$ \\ \hline
\end{tabular}
\label{1}
\end{equation}
with $a\gg c$, $a+c=2b$ (for example $a=99,c=1,b=50$). For future reference
the players moves are labelled $\left| \cdot \cdot \right\rangle $.

It is clear that the unique Nash equilibrium is $\left| 00\right\rangle $,
corresponding to the greedy proposal $\left( a,c\right) $. However, when the
game is played with human players, such greedy proposals are most often
refused, even in one-shot games where the responder has no material or
strategic advantage in refusing the offer. Based on this and similar results
in other situations (public goods games, etc), Bowles and Gintis\cite
{Bowles1} \cite{Bowles2} developed the notion of strong reciprocity (\textit{%
Homo reciprocans}\cite{Bowles4}) as a better model for human behavior. Homo
reciprocans would come to social situations with a propensity to cooperate
and share but would respond to selfish behavior on the part of others by
retaliating , even at a cost to himself and even when he could not expect
any future personal gains from such actions.

Going a step further, the same authors in collaboration with a group of
anthropologists conducted a very interesting ``ultimatum game experiment''
in many small-scale societies around the world\cite{Bowles3}. Consistently
different results are obtained in different societies and the authors
conclude that Homo oeconomicus is rejected in all cases, the players'
behavior being strongly correlated with existing social norms and market
structure in their societies. Apparently human decision problems involve a
mixture of self-interest and a background of (internalized) social norms.

But how does one code for social norms in mathematical games? It is here
that quantum games (or a restricted version thereof) may be of help. In a
full \textit{quantum game}\cite{Meyer} the players have at their disposal an
Hilbert space of strategies rather than a discrete set (or a simplex in the
case of mixed strategies). In practical terms one considers an initial
vector in the tensor product space of moves and then each player can act on
his part of the space by arbitrary unitary operations. In a \textit{%
restricted quantum game} (RQG) version\cite{Eisert}, the initial state is
again an arbitrary vector but the players operations are restricted to
classical moves, that is, to permutations of their basis states.

The restricted version is probably the most appropriate for human decision
problems, because it is not clear how to interpret general unitary
operations in terms of human decisions. On the other hand the choice of the
initial state might be a useful tool to code for the background of social
norms on which classical human moves take place.

For the simplified ultimatum example corresponding to the payoff matrix (\ref
{1}) the Hilbert space is spanned by $\left\{ \left| 00\right\rangle ,\left|
01\right\rangle ,\left| 10\right\rangle ,\left| 11\right\rangle \right\} $,
a general initial state is

\begin{equation}
\phi =c_{00}\left| 00\right\rangle +c_{01}\left| 01\right\rangle
+c_{10}\left| 10\right\rangle +c_{11}\left| 11\right\rangle  \label{2}
\end{equation}
with $\left| c_{00}\right| ^{2}+\left| c_{01}\right| ^{2}+\left|
c_{10}\right| ^{2}+\left| c_{11}\right| ^{2}=1$ and the (classical) players
moves are the matrices $M_{0}=\left( 
\begin{array}{ll}
1 & 0 \\ 
0 & 1
\end{array}
\right) $ and $M_{1}=\left( 
\begin{array}{ll}
1 & 0 \\ 
0 & 1
\end{array}
\right) $ or probabilistic combinations thereof for mixed strategies.

If the initial $\phi $ state is a factorized one, the outcome is the same as
in the classical game but not if it is an entangled state. It is easy to
interpret this effect. It means that when some action is taken (for example
a purely selfish action) in one of the components of the entangled state
that will affect the other components as well and the resulting payoff. 
\textit{Individual decisions have entangled consequences} and the
entanglement is defined by the social environment norms, which in
mathematical terms corresponds to the choice of the $\phi $ state.

With the general initial state in (\ref{2}) the game becomes equivalent to a
three-parameter family of classical games with payoffs

\begin{equation}
\begin{tabular}{|c|c|c|}
\hline
& $R_{0}$ & $R_{1}$ \\ \hline
$P_{0}$ & $a\left| c_{11}\right| ^{2}+b\left| c_{01}\right| ^{2},c\left|
c_{11}\right| ^{2}+b\left| c_{01}\right| ^{2}$ & $a\left| c_{10}\right|
^{2}+b\left| c_{00}\right| ^{2},c\left| c_{10}\right| ^{2}+b\left|
c_{00}\right| ^{2}$ \\ \hline
$P_{1}$ & $b\left| c_{11}\right| ^{2}+a\left| c_{01}\right| ^{2},b\left|
c_{11}\right| ^{2}+c\left| c_{01}\right| ^{2}$ & $b\left| c_{10}\right|
^{2}+a\left| c_{00}\right| ^{2},b\left| c_{10}\right| ^{2}+c\left|
c_{00}\right| ^{2}$ \\ \hline
\end{tabular}
\label{3}
\end{equation}
Each member of the family must have at least a Nash equilibrium in pure or
mixed strategies. Let $\left( \mu ,1-\mu \right) $ and $\left( \nu ,1-\nu
\right) $ be the probabilities for proposer and responder to use moves $%
M_{0} $ and $M_{1}$. Then, their payoffs are, respectively 
\begin{equation}
\begin{array}{lll}
\Bbb{P}_{P} & = & \mu \left( a-b\right) \left\{ \nu \left( \left|
c_{11}\right| ^{2}-\left| c_{01}\right| ^{2}\right) +\left( 1-\nu \right)
\left( \left| c_{10}\right| ^{2}-\left| c_{00}\right| ^{2}\right) \right\}
+f\left( \nu \right) \\ 
\Bbb{P}_{R} & = & \nu \left( \left| c_{11}\right| ^{2}-\left| c_{10}\right|
^{2}\right) \left( \mu c+\left( 1-\mu \right) b\right) +\nu \left( \left|
c_{01}\right| ^{2}-\left| c_{00}\right| ^{2}\right) \left( \mu b+\left(
1-\mu \right) c\right) +g\left( \mu \right)
\end{array}
\label{4}
\end{equation}
with $f\left( \nu \right) =\nu b\left( \left| c_{11}\right| ^{2}-\left|
c_{10}\right| ^{2}\right) +\nu a\left( \left| c_{01}\right| ^{2}-\left|
c_{00}\right| ^{2}\right) +b\left| c_{10}\right| ^{2}+a\left| c_{00}\right|
^{2}$ and $g\left( \mu \right) =\mu \left( c-b\right) \left( \left|
c_{10}\right| ^{2}-\left| c_{00}\right| ^{2}\right) +b\left| c_{10}\right|
^{2}+c\left| c_{00}\right| ^{2}$

From (\ref{3}) one sees that the game is invariant for the replacements $%
\left( c_{11}\leftrightarrow c_{01},c_{10}\leftrightarrow c_{00}\right) $ or 
$\left( c_{11}\leftrightarrow c_{10},c_{01}\leftrightarrow c_{00}\right) $
and has two different classes of Nash equilibria.

(i) If $\left| c_{11}\right| ^{2}-\left| c_{01}\right| ^{2}$ and $\left|
c_{10}\right| ^{2}-\left| c_{00}\right| ^{2}$ are both $\geq 0$ or $\leq 0$
, because of the symmetry one may choose $\left| c_{11}\right| ^{2}\geq
\left| c_{01}\right| ^{2}$ and $\left| c_{10}\right| ^{2}\geq \left|
c_{00}\right| ^{2}$. Then, a Nash equilibrium is obtained for $\mu =1$ and
also for a pure responder strategy $\nu =1$ or $0$ depending on the sign of $%
\nu c\left( \left| c_{11}\right| ^{2}-\left| c_{10}\right| ^{2}\right) +\nu
b\left( \left| c_{01}\right| ^{2}-\left| c_{00}\right| ^{2}\right) $. The
payoff of the responder is the largest of $c\left| c_{11}\right|
^{2}+b\left| c_{01}\right| ^{2}$ or $c\left| c_{10}\right| ^{2}+b\left|
c_{00}\right| ^{2}$ with corresponding proposer payoffs $a\left|
c_{11}\right| ^{2}+b\left| c_{01}\right| ^{2}$ or $a\left| c_{10}\right|
^{2}+b\left| c_{00}\right| ^{2}$.

An example in this class is 
\begin{equation}
\phi =\frac{1}{\sqrt{2}}\left\{ \left| 11\right\rangle +\left|
01\right\rangle \right\}  \label{5}
\end{equation}
for which the Nash equilibrium payoffs are 
\begin{equation}
\begin{array}{lll}
\Bbb{P}_{P} & = & \frac{1}{2}\left( a+b\right) \\ 
\Bbb{P}_{R} & = & \frac{1}{2}\left( c+b\right)
\end{array}
\label{6}
\end{equation}

(ii) If $\left| c_{11}\right| ^{2}-\left| c_{01}\right| ^{2}$ and $\left|
c_{10}\right| ^{2}-\left| c_{00}\right| ^{2}$ have opposite signs, choosing $%
\left| c_{11}\right| ^{2}\geq \left| c_{01}\right| ^{2}$ and $\left|
c_{10}\right| ^{2}\leq \left| c_{00}\right| ^{2}$ the strategy of the
proposer now depends on $\nu $ and Nash equilibria are obtained for pure or
mixed strategies depending on the values of the coefficients. An example is 
\begin{equation}
\phi =\frac{1}{\sqrt{2}}\left\{ \left| 11\right\rangle +\left|
00\right\rangle \right\}  \label{7}
\end{equation}
which has a Nash equilibrium for the mixed strategy $\mu =\nu =\frac{1}{2}$
and payoffs 
\[
\begin{array}{lll}
\Bbb{P}_{P} & = & \frac{1}{4}\left( a+b\right) \\ 
\Bbb{P}_{R} & = & \frac{1}{4}\left( c+b\right)
\end{array}
\]

A even wider range of possibilities and payoff structures may be simply
obtained by increasing the number of possible proposer moves in the original
payoff matrix.

In conclusion: One sees that the restricted quantum game (RQG) structure,
while keeping the rational self-interest choice characteristic of the Nash
equilibria, does so in a background that allows for the coding of social
norms. This occurs because of the entangled nature of the $\phi $ state.

As shown, a RQG is equivalent to a family of classical games. Therefore one
might code social norms (as well as player contracts) directly on the family
of classical games without any reference to quantum games. Nevertheless the
coding of non-trivial members of the family by a simple choice of an
entangled $\phi $ vector seems to be an useful compact way to characterize
such families.


\begin{thebibliography}{9}
\bibitem{Guth}  W. G\"{u}th, R. Schmittberger and B. Schwarze; \textit{An
experimental analysis of ultimatum bargaining}, J. of Economic Behavior and
Organization, 3 (1982) 367-388.

\bibitem{Bowles1}  S. Bowles and H. Gintis; \textit{The evolution of strong
reciprocity}, Santa Fe Institute Working paper 98-08-073.

\bibitem{Bowles2}  S. Bowles and H. Gintis; \textit{The evolution of
reciprocal preferences}, Santa Fe Institute Working paper 00-12-072.

\bibitem{Bowles4}  S. Bowles and H. Gintis; \textit{Homo reciprocans, }%
Nature 415 (2002) 125-127.

\bibitem{Bowles3}  J. Heinrich, R. Boyd, S. Bowles, C. Camerer, E. Fehr, H.
Gintis and R. McElreath; \textit{Cooperation, reciprocity and punishment in
fifteen small-scale societies}, American Economics Review 91 (2001) 73-78.

\bibitem{Meyer}  D. A. Meyer; \textit{Quantum strategies}, Phys. Rev.
Letters 82 (1999) 1052-1055.

\bibitem{Eisert}  J. Eisert, M. Wilkens and M. Lewenstein; \textit{Quantum
games and quantum strategies}, Phys. Rev. Letters 83 (1999) 3077-3080.
\end{thebibliography}
\end{document}